\documentclass{main}

\def\be{\begin{equation}}
\def\ee{\end{equation}}
\def\bea{\begin{eqnarray}}
\def\eea{\end{eqnarray}}

\newcommand{\Photo}{}

\usepackage{xspace}
\usepackage{lineno}
\usepackage{cite}

\newcommand{\KK}            {\ensuremath{\rm{K}^{+}\rm{K}^{-}}\xspace}
\newcommand{\pipi}            {\ensuremath{\pi^{+}\pi^{-}}\xspace}
\newcommand{\snn}          {\ensuremath{\sqrt{s_{\mathrm{NN}}}}\xspace}
\newcommand{\PbPb}         {\mbox{Pb--Pb}\xspace}
\newcommand{\ptsquared}    {\ensuremath{p_{\rm{T}}^{2}}\xspace}
\newcommand{\MeVmass} {Me\kern-.1emV/$c^2$\xspace}
\newcommand{\GeVmass} {Ge\kern-.1emV/$c^2$\xspace}
\newcommand{\GeVcSquared} {\ensuremath{(\mathrm{Ge\kern-.1emV}/c)^{2}}\ }
\DeclareUnicodeCharacter{2212}{-}
\begin{document}

\title{\Large $\mathrm{K}^{+}\mathrm {K}^{−}$ photoproduction in ultra-peripheral Pb--Pb collisions}

\author{Minjung Kim for the ALICE Collaboration}

\address{Department of Physics, University of California, Berkeley, CA, USA \\ Center for Frontiers in Nuclear Science, Stony Brook University, Stony Brook, NY, USA}

\maketitle\abstracts{
In ultra-peripheral collisions (UPCs) of relativistic heavy ions, photoproduction occurs when a photon emitted from one nucleus interacts with the other nucleus from the opposing beam, producing particles in the final state. Measurements of \KK photoproduction probe interactions and couplings between the $\phi (1020)$ and charged kaons with photons and nuclear targets. We report exclusive \KK photoproduction cross section at midrapidity in \PbPb collisions at \snn = 5.02 TeV, which is measured for the first time in UPCs.
}


\section{Introduction}
In ultra-relativistic heavy-ion collisions, the electromagnetic field of a fast-moving heavy ion acts as a source of quasi-real photons~\cite{Bertulani:2005ru,Baltz:2007kq}. Those emitted photons can interact with the nucleus from the opposing beam. Ultra-peripheral collisions (UPCs) are those in which the impact parameter between the two colliding nuclei is greater than the sum of their nuclear radii. Photonuclear interactions can be effectively studied in UPCs since there are no hadron-hadron interactions present.

One of the interesting processes of photonuclear interactions is exclusive production of a vector meson. A virtual photon fluctuates into a quark-antiquark pair and elastically scatters off the nucleus, emerging as a vector meson~\cite{Klein:2019qfb}. The exclusivity of the process necessitates that there be no net colour charge transfer to the target, and therefore requires the exchange of at least two gluons with the target. Experimentally, the process can be rather easily identified by a rapidity gap between the produced vector meson and the target. Light vector mesons are typically reconstructed via their decay into oppositely charged meson pair, e.g., $\rho^{0}(700) \rightarrow \pi^+\pi^-$ or $\phi(1020) \rightarrow \rm{K}^+\rm{K}^-$.

However, oppositely charged meson pairs, \pipi or \KK, can emerge from different photonuclear interactions, where the photon fluctuates directly into a virtual meson pair~\cite{Klein:1999gv}. One of the mesons can then scatter elastically from the target, making the pair real. Both processes contribute to the cross section measurements of \pipi or \KK pairs, making their distinction impractical. In order to further separate these two components, the production amplitude is described with two terms: the resonance with amplitude $A_{\phi}$, and the continuum with amplitude $B_{\rm{KK}}$, giving~\cite{Soding:1965nh}

\begin{equation}
\frac{\mathrm{d}\sigma}{\mathrm{d}M_{\rm{KK}}} = \bigg| A_{\phi}
\frac{\sqrt{M_{\rm{KK}}M_{\phi}\Gamma_\phi}}
{M_{\rm{KK}}^2-M_{\phi}^2+iM_\phi\Gamma_\phi}+B_{\rm{KK}}\bigg|^2,
\label{eq:BW}
\end{equation}
where $M_{\phi}= 1019.416 \pm 0.016$ \MeVmass~\cite{ParticleDataGroup:2022pth} and $\Gamma_{\phi}$ are the $\phi (1020)$ mass and mass-dependent width, respectively, with
\begin{equation}
    \Gamma_{\phi}= \Gamma_0 \frac{M_{\phi}}{M_{\rm{KK}}} 
    \bigg(\frac{M_{\rm{KK}}^2-4M_{\rm{K}}^2}{M_\phi^2-4M_{\rm{K}}^2}\bigg)^{3/2}.
\end{equation}
Here $\Gamma_0 = 4.249 \pm 0.013$ \MeVmass is the native \mbox{$\phi$(1020)} width~\cite{ParticleDataGroup:2022pth}. $M_{\rm{K}} = 493.677 \pm 0.016$ \MeVmass is the kaon mass~\cite{ParticleDataGroup:2022pth}. By taking $A_{\phi}$ to be real, the relative phase between $\phi(1020) \rightarrow \rm{K}^+\rm{K}^-$ and direct \KK continuum is encoded in $B_{\rm{KK}}$.

This article describes the first cross section measurement of the final state \KK in exclusive photoproduction in UPCs with the ALICE detector~\cite{ALICE:2023kgv}. The measurement covers the invariant mass of \KK from 1.1 to 1.4 \GeVmass, above the $\phi(1020)$ meson peak, with the lower mass limit set at about $M_{\phi} + 18\Gamma_{0}$. The results are presented as functions of $M_\mathrm{KK}$ and the dikaon transverse momentum ($p_{\rm{T,KK}}$) in the rapidity range $|y_\mathrm{KK}|\ <\ 0.8$.
Additionally, studies regarding the significance of the resonance contribution and its interference with the direct production are also reported.

\section{ALICE detector, data and analysis method}
ALICE (A Large Ion Collider Experiment) is a general-purpose detector located at the LHC~\cite{ALICE:2008ngc,ALICE:2014sbx}. In  the central barrel of ALICE, the Time Projection Chamber (TPC)~\cite{Alme:2010ke} is used for charged particles tracking, as well as particle species identification based on the specific energy loss (d$E$/d$x$). The Inner Tracking System (ITS)~\cite{ALICE:2010tia} consisting of 6 layers of silicon detectors close to the beam pipe complements the TPC for tracking. In addition, the two innermost layers of the ITS are composed with silicon pixel detectors (SPD), contributes to the trigger decision. The data for this analysis was collected by the ALICE detector during the LHC Run 2 in 2015. The Time Of Flight (TOF) detector together with SPDs triggered events having two back-to-back tracks in the transverse plane in central barrel. At the same time, the events were vetoed by the V0~\cite{ALICE:2013axi} and AD scintillators placed along the beam pipe, covering large pseudorapidity range~\cite{LHCForwardPhysicsWorkingGroup:2016ote,Broz:2020ejr} to ensure exclusivity~\cite{ALICE:2020ugp}. The integrated luminosity of the data sample used in this analysis corresponds to 0.406 $\mu\rm{b}^{-1}$ with 2.6\% uncertainty.

\begin{figure}[h]
	\centering 
	\includegraphics[width=0.9\textwidth]{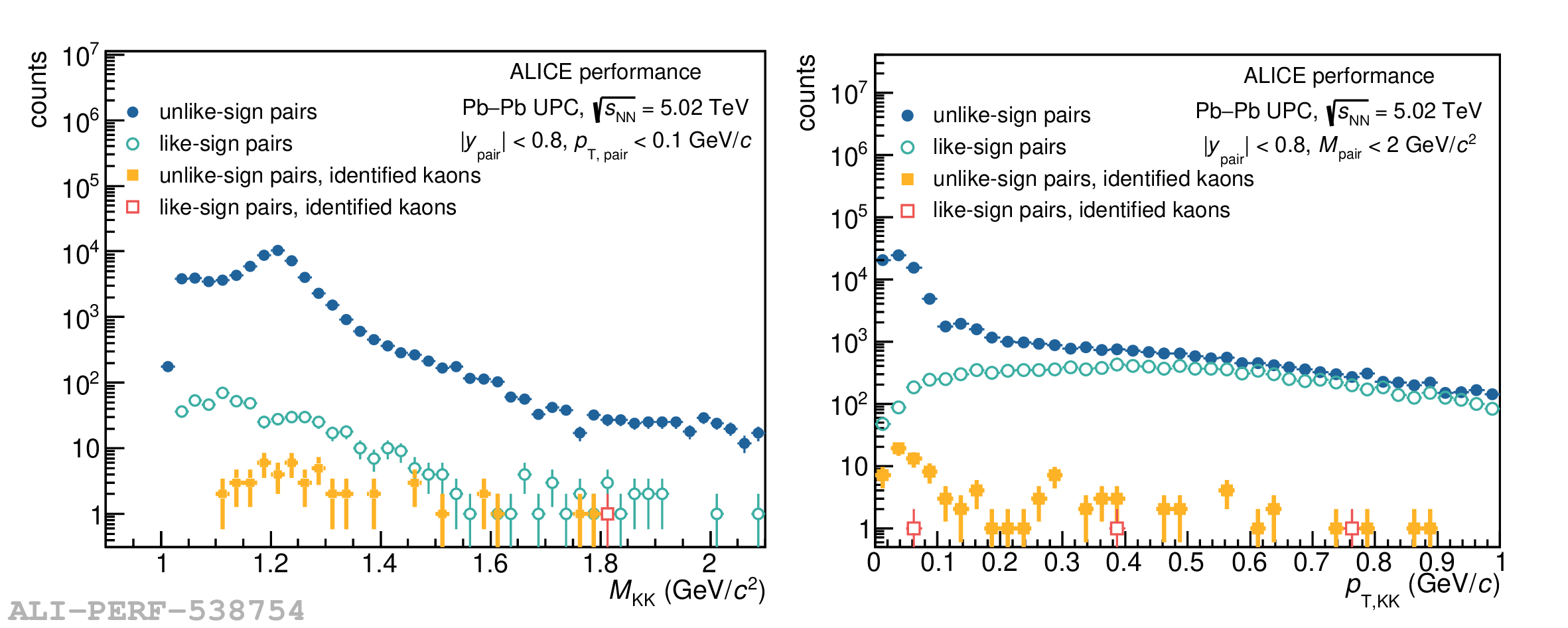}	
	\caption{Raw invariant mass (left) and transverse momentum (right) distributions for opposite-sign and same-sign pairs of inclusive charged particles and identified kaons. Identified kaons refer to satisfying kaon selection criteria based on measured TPC d$E$/d$x$ described in the text. } 
	\label{fig_rawdist}%
\end{figure}

Two oppositely charged tracks form a pair candidate. Figure~\ref{fig_rawdist} shows the kinematic distributions of dikaon pairs as functions of $M_{\rm{KK}}$ and $p_{\rm{T,KK}}$ for opposite and same-charge pairs. By comparing the absolute yield of opposite-sign pairs to same-sign pairs, combinatorial background is found to be negligible in this data set. A rise of candidate opposite-sign pair yield at low $p_{\rm{T,KK}}$, similar to previously observed $\rho^{0}(700) \rightarrow \pi^+\pi^-$ in ALICE \cite{ALICE:2020ugp}, as well as in STAR \cite{STAR:2017enh}, indicates that pair candidates mostly originate from coherent photoproduction. 

\begin{figure}[h]
	\centering 
	\includegraphics[width=0.47\textwidth]{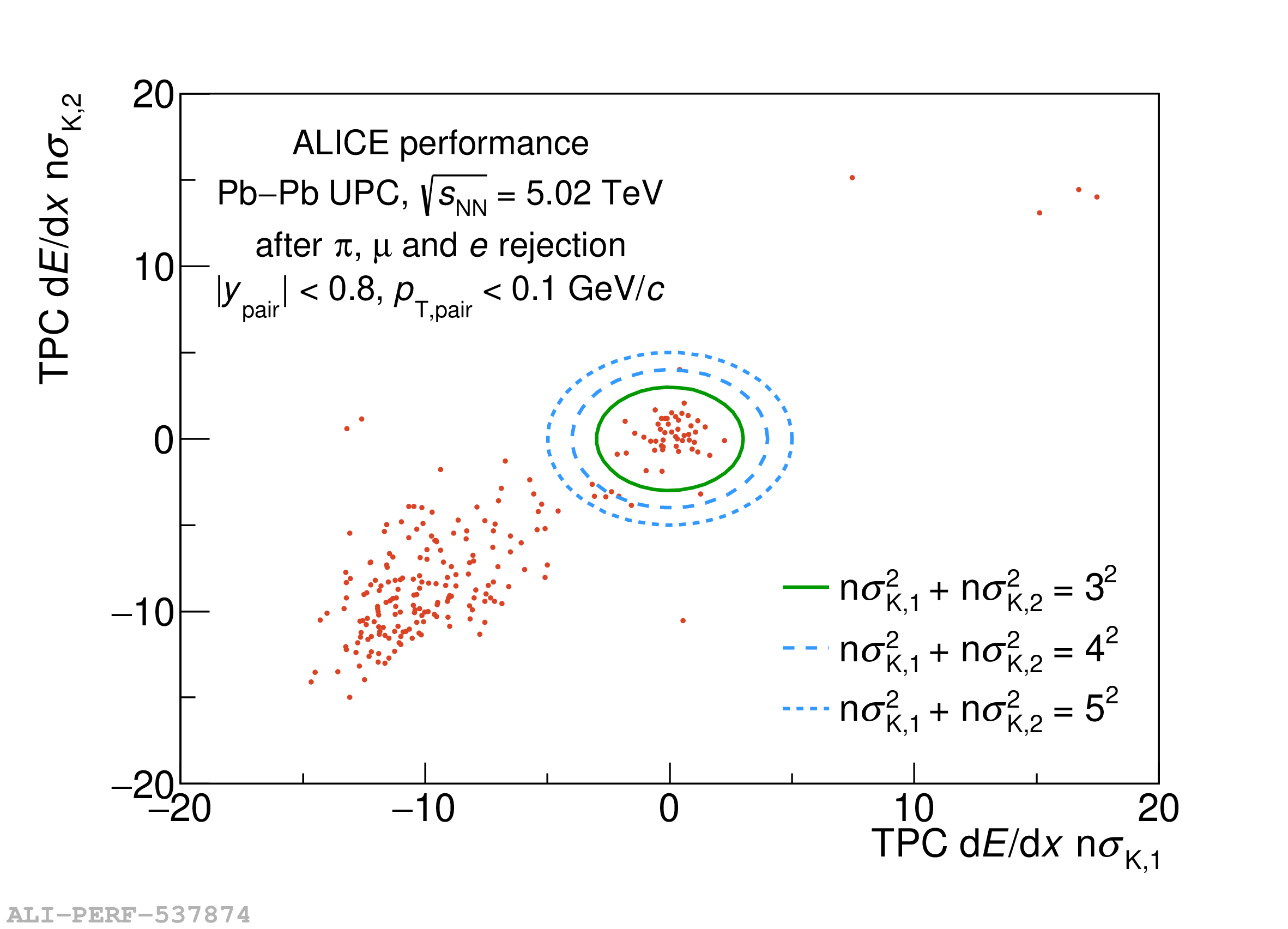}	
	\includegraphics[width=0.47\textwidth]{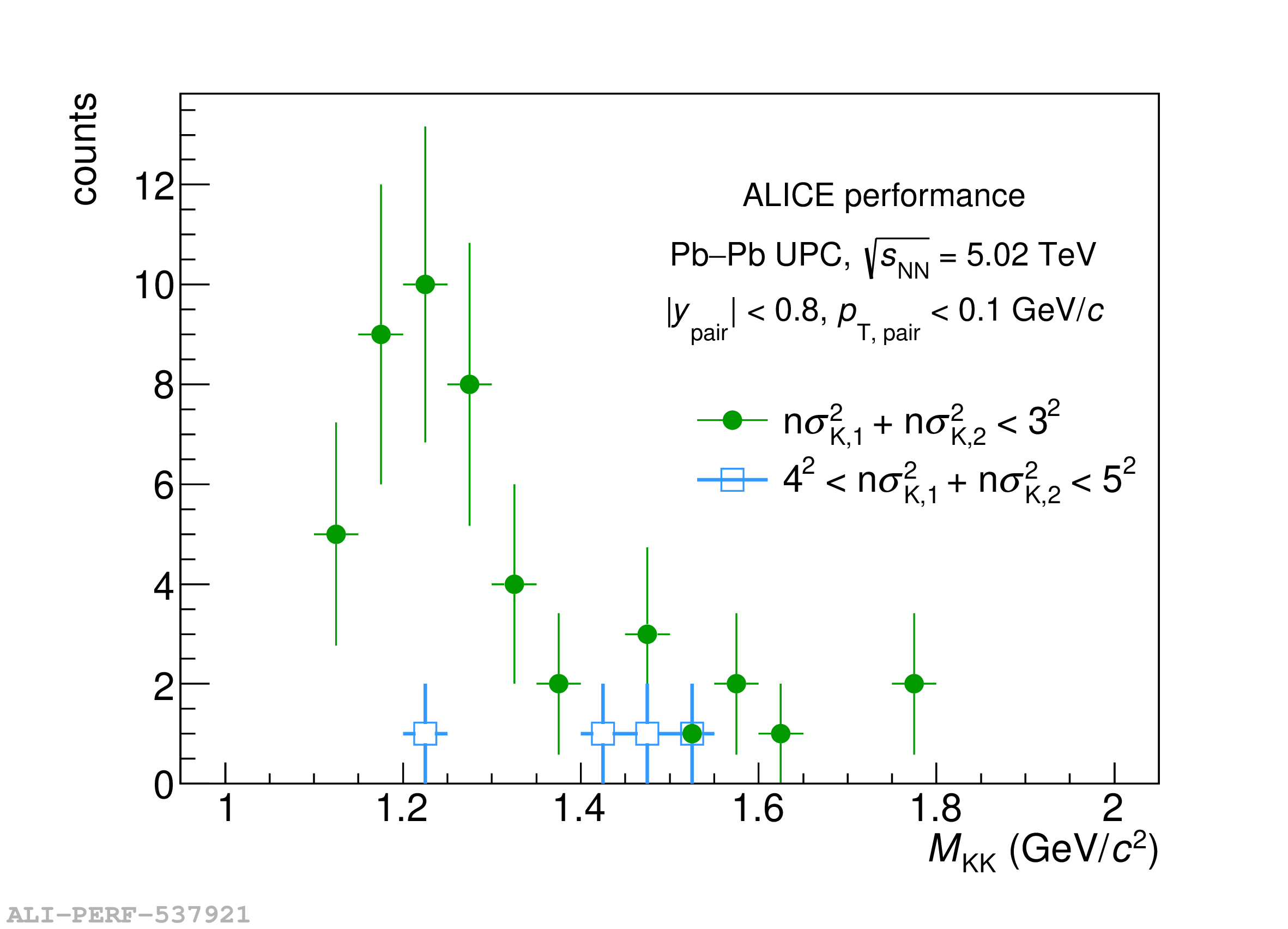}	
	\caption{Left: TPC d$E$/d$x$ correlation between positive and negative charged kaon candidates in each pair represented as the deviation of the measured TPC d$E$/d$x$ from the expected signal in units of the d$E$/d$x$ measurement uncertainty for kaon mass hypothesis. Signal region is defined as $n_{\sigma_{\rm{K}}}^2 + n_{\sigma_{\rm{K}}}^2 < 3^2$ while the region used for background estimation is $4^{2}< n_{\sigma_{\rm{K}}}^2 + n_{\sigma_{\rm{K}}}^2 < 5^2$. Circular boundaries around (0,0) with different radii are presented. Right: Raw number of kaon candidates in the signal region projected into pair invariant mass ($\it{ M_{\rm{KK}}}$) presented in green solid circle. The projection of the estimated background is shown in blue open square.} 
	\label{fig_pid}%
\end{figure}

In order to reject the background contributions from \pipi and dilepton pairs, stringent kaon selection criteria are required based on TPC d$E$/d$x$ information. The selection criteria are applied to the variable $n_{\sigma_{i}}$, the deviation of the measured signal from the expected signal in units of the d$E$/d$x$ measurement uncertainty for each particle hypothesis $i$, where $i$ are $\pi, \mu, e, \mathrm{K}$. The tracks compatible within $2 n_{\sigma_{\pi, \mu, e}}$ are excluded for further analysis. The TPC d$E$/d$x$ $n_{\sigma_{\mathrm{K}}}$ correlation between the two kaon candidates is presented in the left panel in Fig.~\ref{fig_pid}. The remaining background is clustered around (-5,-5) while signal kaon pair candidates are centered around (0,0). Possible misidentified pair candidates within $|n_{\sigma_{\rm{K}}}| < $  3 are estimated using different $n_{\sigma_{\rm{K}}}$ boundaries shown as circular contours in the left panel in Fig.~\ref{fig_pid}. The corresponding yields of signal region, $n_{\sigma_{\rm{K}}}^2 + n_{\sigma_{\rm{K}}}^2 < 3^2$, and background region, $4^{2}< n_{\sigma_{\rm{K}}}^2 + n_{\sigma_{\rm{K}}}^2 < 5^2$, are projected as a function of pair invariant mass in the right panel in Fig.~\ref{fig_pid}. The yield in the background region with respect to the one in the signal region is tiny in the range of 1.1 to 1.4 \GeVmass. Thus, misidentified background is estimated to be negligible in the signal region $|n_{\sigma_{\rm{K}}}| < $  3 in the range of 1.1 to 1.4 \GeVmass.

\section{Cross section results}
The cross section of exclusive \KK photoproduction is obtained by the number of \KK candidates, corrected by acceptance and efficiency, and the integrated luminosity. The acceptance and efficiency values are derived from Monte Carlo simulation with STARLight~\cite{Klein:2016yzr}, and then complemented by a full simulation of the ALICE detector using GEANT3~\cite{Brun:1994aa} to model a realistic detector response. 
\subsection{\ptsquared-differential cross section for exclusive \KK photoproduction}
\begin{figure}[h]
	\centering 
	\includegraphics[width=0.55\textwidth]{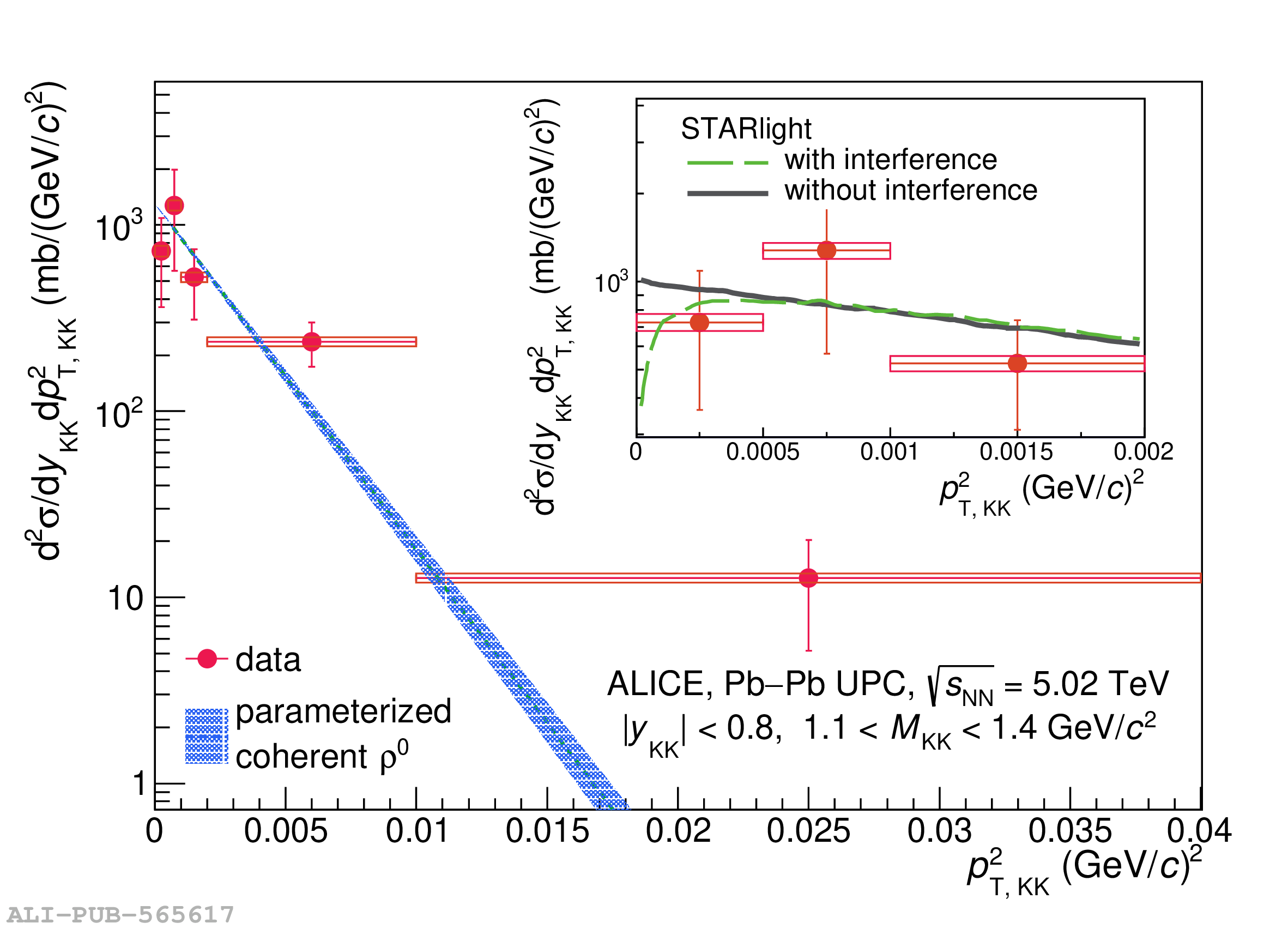}
	\caption{\ptsquared-differential cross section for exclusive \KK photoproduction in $|y_{\rm{KK}}|<0.8$ in Pb--Pb UPCs at \mbox{$\sqrt{s_{\rm{NN}}}$= 5.02 TeV} with parameterized cross section  $\rm{d}^2\it{\sigma}/\rm{d}\it{y}\rm{d}\it{p}_{\rm{T}}^{\rm{2}}$ = $a\rm{exp(-}\it{b}\ptsquared\rm{)}$, where the slope parameter $b$ is fixed to $b$ = 428 $\pm$ 6 (stat.) $\pm$ 15 (syst.) (GeV/$c$)$^{−2}$, taken from previously measured coherent $\rho^{0}$ photoproduction~\cite{ALICE:2015nbw}. Figure from Ref.~\cite{ALICE:2023kgv}.}
	\label{fig_pt}%
\end{figure}

The \ptsquared-differential cross section is shown in Fig.~\ref{fig_pt}. The majority of the cross section is found below $\it{p}_{\rm{T, KK}}^{\rm{2}} <$ 0.01 $(\rm{GeV}/c)^{2}$, consistently with coherent photoproduction. The cross section for the coherent region are well described with an exponential shape $\rm{d}^2\it{\sigma}/\rm{d}\it{y}\rm{d}\it{p}_{\rm{T}}^{\rm{2}}$ = $a\rm{exp(-}\it{b}\ptsquared\rm{)}$, where the slope parameter $b$ is fixed to that measured for coherent $\rho^{0}$ photoproduction on Pb nuclei, $b$ = 428 $\pm$ 6 (stat.) $\pm$ 15 (syst.) (GeV/$c$)$^{−2}$~\cite{ALICE:2015nbw}. 

\subsection{Invariant mass distribution of coherent \KK photoproduction}
\begin{figure}[h]
	\centering 
	\includegraphics[width=0.485\textwidth]{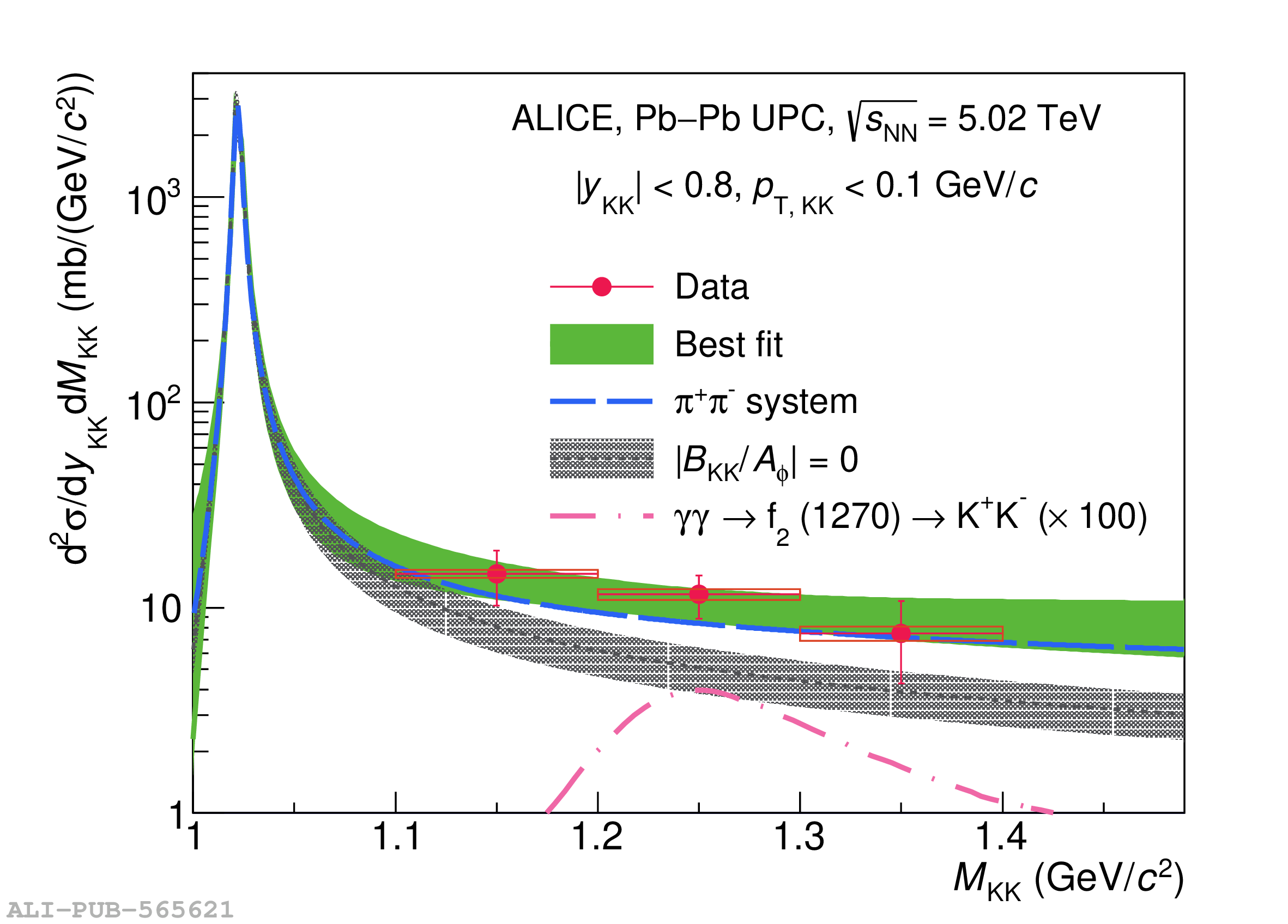}
	\includegraphics[width=0.485\textwidth]{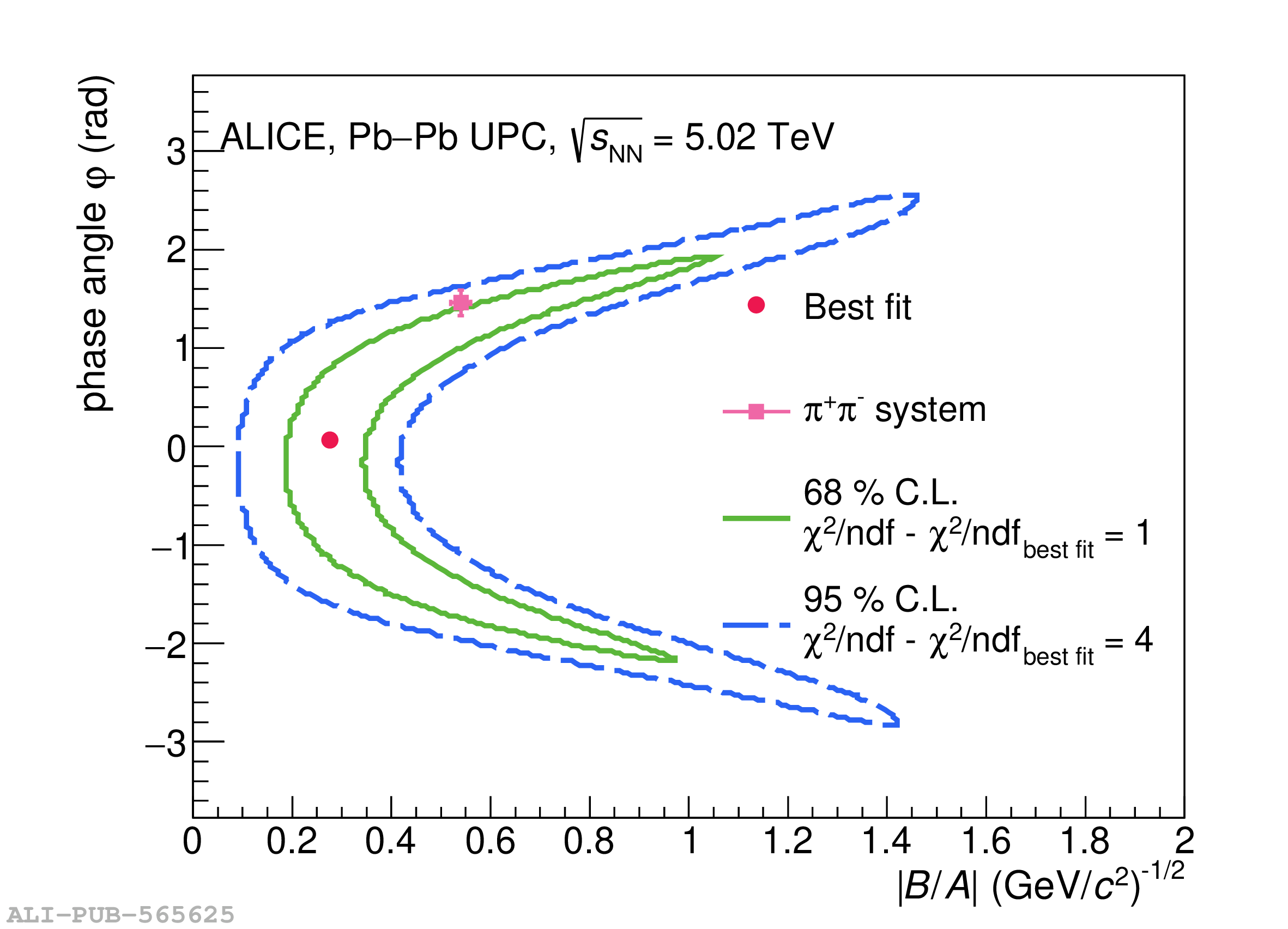}
	\caption{Left: Differential cross section of coherent \KK photoproduction as a function $\it{M_\mathrm{KK}}$ in Pb--Pb UPCs at \mbox{\snn = 5.02 TeV} in $|y_\mathrm{KK}|\ <\ 0.8$. The lines and boxes along the data points represent statistical and systematic uncertainties, respectively. Best fit using Eq.~\ref{eq:BW} (green solid band), parameterized cross section based on $\rho^0$ and direct $\pi^{+}\pi^{-}$ production (blue long-dashed line) and estimated cross section without direct \KK ($\varphi$) contribution (grey plaid band) are shown. $\gamma\gamma \rightarrow f_{2}$(1270)$ \rightarrow \KK$ (pink dashed-dotted line) is factored by 100 for better visualization.   Right: Confidence regions for the relative fraction of direct \KK contribution with respect to the amplitude of $\phi (1020) \rightarrow$ \KK ($|B_{\rm{KK}}/A_{\phi}|$) and the relative phase angle between $\phi(1020)\rightarrow$\KK and direct \KK ($\varphi$). The green solid line and blue dashed line represent the boundary of 68\% and 95\% confidence regions, respectively. Figure from Ref.~\cite{ALICE:2023kgv}. } 
	\label{fig_mass}%
\end{figure}

The left panel in Fig.~\ref{fig_mass} shows the invariant mass distribution for coherent \KK photoproduction as a function of dikaon mass while rejecting most of the incoherent photoproduction contribution by requiring $\ptsquared > $0.01 (GeV/$c$)$^{2}$. 
The spectrum is shown together with the best fit using Eq.~\ref{eq:BW} and also compared to various sets of relative fraction of direct \KK contribution with respect to the amplitude of $\phi (1020) \rightarrow$ \KK ($|B_{\rm{KK}}/A_{\phi}|$) and the relative phase angle between $\phi(1020)\rightarrow$\KK and direct \KK ($\varphi$). Calculations using STARLight for \KK production via reactions such as $\gamma\gamma \rightarrow f_{2}$(1270)$ \rightarrow \KK$ (pink dashed-dotted line) show that these contributions are negligible. 
The black dotted line and the shaded region indicate the results under the assumption of no direct \KK production. The calculations underestimate the measured cross section by about 2.1 $\sigma$ in the range 1.1 $< M_{\rm{\KK}} <$ 1.4 GeV/$c^{2}$. The blue dashed line represents the prediction based on fixed values for $B_{\pipi}/A_{\rho}$~\cite{STAR:2017enh} and relative phase angle $\varphi$~\cite{ALICE:2020ugp}, determined from the measurement of the final \pipi state system of $\rho^{0}$ meson decay and direct \pipi production. This prediction is in agreement with the measured data points, showing slightly lower values yet still within the experimental uncertainties.

The right panel of Fig.~\ref{fig_mass} shows the confidence region for $|B_{\rm{KK}}/A_{\phi}|$ and $\varphi$. A strong correlation between the two parameters produces a horseshoe-shaped distribution in their phase space.
The solid green and dashed blue band indicate the 68$\%$ and 95$\%$ confidence level, respectively. The best-fit point shows that the results of the \KK measurement is fully compatible with the prediction from the \pipi system.

\section{Conclusion}
We report the first study of coherent \KK photoproduction in ultra-peripheral collisions at the centre-of-mass energy per nucleon of the photon–nucleus15
(Pb) system $W_{\gamma \mathrm{Pb, n}}$ from
33 to 188 GeV, in the range $1.1 < M_{\rm{KK}} < 1.4$ \GeVmass and $|y_{\mathrm{KK}}|<0.8$. The measured cross section is concentrated below \ptsquared $ < 0.01 \ $\GeVcSquared, consistent with coherent photoproduction.
The measured cross section is about 2.1 $\sigma$ larger than what is expected only from $\phi (1020)$ production, but is consistent with a mixture of $\phi (1020)$ and direct \KK production.  The fitted ratio of $\phi (1020)$ production to $\KK$ production is consistent with that seen for the $\rho^0$ and direct $\pi^{+}\pi^{-}$ production.

\end{document}